\documentclass{pnastwo}

\usepackage{amsmath,amssymb,graphicx}

\usepackage{amssymb,amsfonts,amsmath}

\newcommand{\rearth}{$R_\oplus$}

\newcommand{\mearth}{$M_\oplus$}

\newcommand{\rsun}{$R_\odot$}

\newcommand{\rstar}{$R_*$}

\newcommand{\rpl}{$R_{\rm P}$}

\newcommand{\apjs}{ApJS}
\newcommand{\mnras}{MNRAS}
\newcommand{\apj}{ApJ}
\newcommand{\aap}{A\&A}
\newcommand{\apjl}{ApJL}

\newcommand{\nat}{Nature}
\newcommand{\aj}{AJ}
\newcommand{\araa}{ARAA}

\contributor{Published in Proceedings
of the National Academy of Sciences of the United States of America}
\url{www.pnas.org/cgi/doi/10.1073/pnas.0709640104}
\copyrightyear{2008}
\issuedate{September 2, 2014}
\volume{111}
\issuenumber{35}

\begin{document}



\title{Exploring Exoplanet Populations with NASA's Kepler Mission}





\author{Natalie M. Batalha\affil{1}{NASA Ames Research Center, Moffett Field, CA, USA}}

\contributor{Submitted to Proceedings of the National Academy of Sciences of the United States of America}

\maketitle

\begin{article}

\begin{abstract} 
The Kepler Mission is exploring the diversity of planets and planetary systems.  Its legacy will be a catalog of discoveries sufficient for computing planet
occurrence rates as a function of size, orbital period, star-type, and insolation flux. The mission has made significant progress toward achieving that goal.  Over 3\,500 
transiting exoplanets have been identified from the analysis of the first three years of data, 100 of which are in the habitable zone.  The catalog has a high reliability rate (85-90\% averaged over the period/radius plane) which is improving as follow-up observations continue.  Dynamical (e.g. velocimetry and transit timing) and statistical methods have confirmed and characterized hundreds of planets over a large range of sizes and compositions for both single and multiple-star systems.  Population studies suggest that planets abound in our galaxy and that small planets are particularly frequent.  Here, I report on the progress Kepler has made measuring the prevalence of exoplanets orbiting within 1 AU of their host stars in support of NASA's long-term goal of finding habitable environments beyond the solar system.
\end{abstract}

\section{Significance Statement}
Kepler is NASA's first mission capable of detecting Earth-size planets.  Four years of spacecraft data have yielded thousands of planet discoveries and have lifted our blinders to the small planets that populate the galaxy.  The mission has enabled studies of exoplanet populations so critical for future efforts to find habitable environments and evidence of life beyond the solar system.







\section{NASA's 10th Discovery Mission}

Searching for evidence of life beyond Earth is one of the primary goals of science agencies in the US and abroad. The goal looms closer as a result of exoplanet discoveries made by NASA's 10th Discovery mission, Kepler.  Launched in March 2009, the Kepler spacecraft is exploring the diversity of planets and planetary systems within 1 AU.  The primary mission objective is to determine the prevalence of potentially habitable, earth-size planets in the galaxy.  Discovering exo-terrans in the habitable zone, characterizing those that have habitable environments and then focusing on the signatures of biological chemistry is a path of exploration that stretches decades into the future.  It begins by determining if planets like Earth are abundant.

From 2009 to 2013, Kepler monitored a 115 square degree field in the constellations Cygnus and Lyra, collecting ultra-high precision photometry of over 190,000 stars  simultaneously at a 30-minute cadence.  Nearly uninterrupted photometry is possible due to a heliocentric orbit and off-ecliptic pointing.  The observations yield an evenly-sampled, minimally-gapped flux time series  that can be searched for periodic diminutions of light due to the transit of an exoplanet across the stellar disk in an aligned geometry.  The photometer was engineered to achieve 20 ppm relative precision in 6.5 hours for a 12th magnitude G-type main sequence star \cite{koch10}.  For reference, the Earth orbiting the Sun would produce an 84 ppm signal lasting approximately 13 hours.

Kepler's pixel and flux measurements \cite{jenkins_data} are publicly available at the Mikulski Archive for Space Telescopes\footnote{\url{http://archive.stsci.edu/kepler}} (MAST).  Transit searches have been performed on successively larger data volumes yielding incremental planet candidate catalogs that are hosted at NASA's Exoplanet Archive\footnote{\url{http://exoplanetarchive.ipac.caltech.edu/}} (NEA). To date, approximately three-quarters of the data have been thoroughly searched. As of this writing (April 2014), the archive is host to over 3500 viable planet candidates (with radius smaller than twice Jupiter).  All have been subjected to a series of statistical tests (based on the Kepler data itself) that ensure a low rate of instrumental and astrophysical false positives \cite{wuDV}.

Kepler has a follow-up observation program (FOP) to increase the reliability of the catalog even further by a) improving the accuracy of the host star properties which in turn improves the accuracy of the planet properties (or changes the interpretation altogether) and b) identifying bound stellar companions and line-of-sight neighbors that might indicate an astrophysical false positive.  Ground-based and space-based telescopes with apertures ranging from 1.5 to 10 meters are being employed to acquire high resolution spectroscopy and high-contrast/high spatial resolution images.  Strategic high-precision Doppler measurements are providing planet masses in an effort to delineate the transition between terrestrial and giant planets.

Translating Kepler's discovery catalog into population statistics requires corrections for observation and detection biases.  This is a work in progress.  However, occurrence rate calculations based on subsets of the data already indicate that nature produces small planets relatively efficiently in the warmer environs of a planetary system.  Giant planets in such orbits are orders of magnitude less frequent than their sub-Neptunian counterparts.  Ironically, the hot-Jupiters that comprised the very first Doppler and transiting exoplanet discoveries are actually quite rare.  Current results for habitable zone planets tells us that we may not have to look very far before happening upon a planet similar to Earth.  

A comprehensive review of Kepler exoplanet science is beyond the scope of this contribution.  Here, I focus on the science leading to the determination of planet occurrence rates, from the discovery catalogs to the first calculations of the prevalence of earth-like planets.

\section{Kepler Transforms the Discovery Space}

Exoplanet discoveries trickled in at a steady rate in the latter half of the 90's.   Approximately thirty were reported with sizes ranging from 0.4 to 8 jupiter masses and orbital periods ranging from 3 to 3800 days.  Heralding in the new millennium, the first transiting exoplanet was discovered  \cite{hd209458_henry,hd209458_ch}.  The timing was a boon for Kepler as it was proposing to utilize this detection technique from space.  In 2000, Kepler was one of the three Discovery mission proposals invited to submit a Concept Study Report.  It was selected for flight on December 20, 2001.

\begin{figure*}
\includegraphics[width=1.0\textwidth]{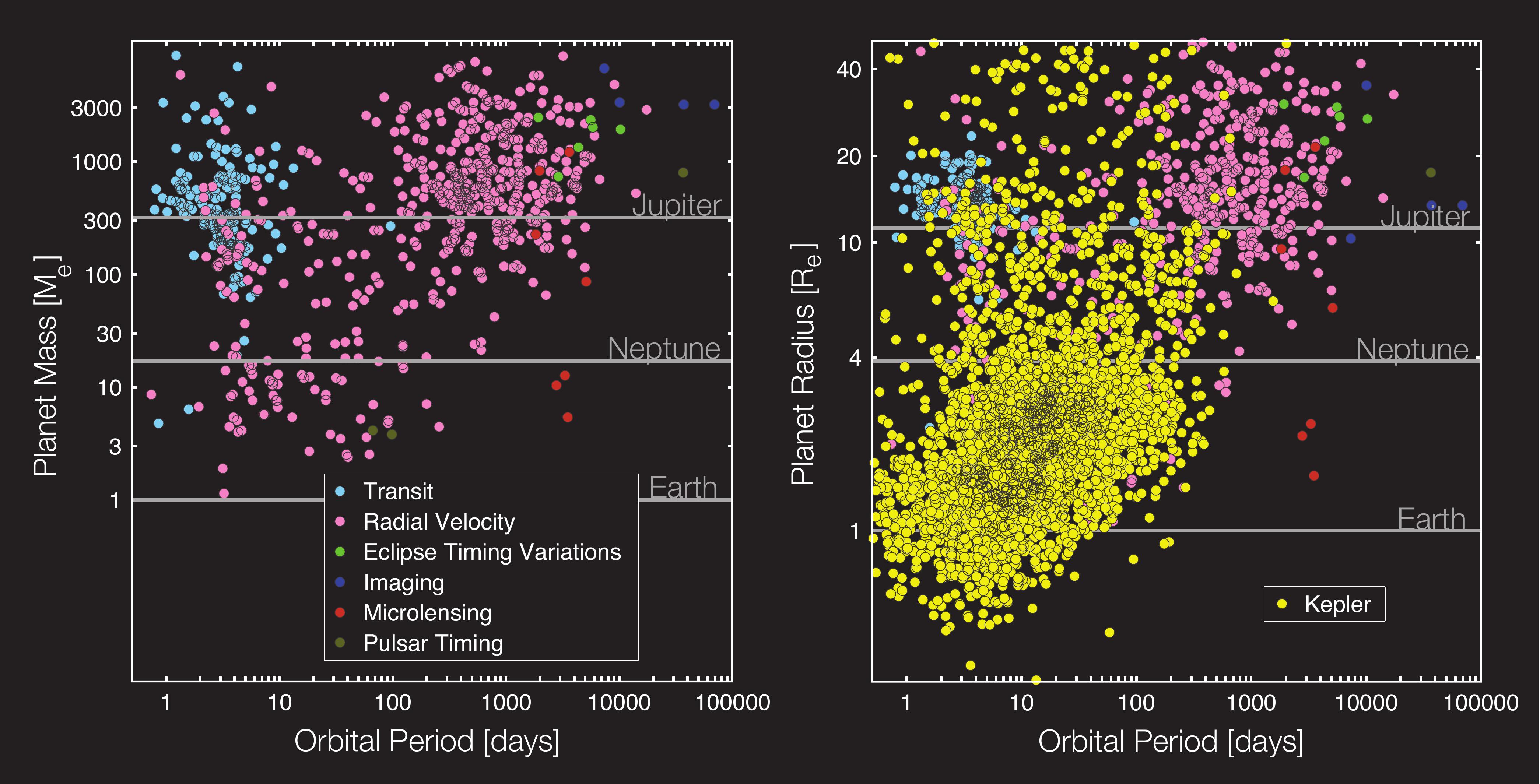}
\caption{Non-Kepler exoplanet discoveries (left) are plotted as mass versus orbital period, colored according to the detection technique.  A simplified mass-radius relation is used to transform planetary mass to radius (right), and the $>3500$ Kepler discoveries (yellow) are added for comparison.  86\% of the non-Kepler discoveries are larger than Neptune while the inverse is true of the Kepler discoveries:  85\% are smaller than Neptune.}
\label{fig:discoveries}
\end{figure*}

As Kepler was being designed and built, exoplanet discoveries were growing at an accelerated pace.  By the eve of Kepler's launch, over 300 discoveries had been reported including nearly 70 transiting systems.  All non-Kepler discoveries up through April 2014 are shown in Figure~\ref{fig:discoveries} (left panel), in a plot of mass (or minimum mass for non-transiting planets) versus orbital period with symbols color coded by the discovery method.  (Methodologies with small numbers of discoveries have been left out for clarity).  Collectively, there are 697 (non-Kepler) exoplanets (with a measured orbital period and radius or mass) associated with 583 unique stars.  Approximately 16\% of these host stars are known to harbor multiple planets.  

The right panel of Figure~\ref{fig:discoveries} shows the same population together with the Kepler planet candidate discoveries in the cumulative table at NEA as of April 2014.  Detections are plotted as planet radius versus orbital period, and the non-Kepler discoveries are included for comparison.  Where planet radii are not available (as is the case for most of the Doppler detections), they are estimated using a polynomial fit to solar system planets ($R = M^{0.4854}$) \cite{architecture1}.  Shown here are 3,553 Kepler discoveries associated with 2,658 stars.  Approximately 22\% of the Kepler host stars are known to harbor multiple planet candidates.  The overall reliability of the catalog (80 - 90\%) is discussed below.  

The demographics of the observed population has changed remarkably.  Kepler has increased the roster of exoplanets by nearly 400\%.  More remarkable still is the change in the distribution: 86\% of the non-Kepler discoveries have {\em masses} larger than Neptune whereas 85\% of the Kepler discoveries have {\em radii} smaller than Neptune.  Kepler is filling in an area of parameters space that wasn't previously accessible.  The increase in sensitivity afforded us by Kepler has opened the floodgates to the small planets so difficult to detect from ground-based surveys.  The most common type of planet known to us is a population that doesn't exist in our own Solar System: the super-earths and mini-neptunes between 1 and 4 earth-radii.

 \section{Status of Kepler's Discovery Catalogs}\label{sec:candidates}

Catalogs of Kepler's viable planet candidates have been released periodically since launch and have included 312, 1235, 2338, 2738, and 3553 detections (cumulative counts) associated with 306, 997, 1797, 2017, and 2658 stars based on 1.5, 13, 16, 22, and 34.5 of the $\sim 48$ months of data acquired during the primary mission \cite{june2010Catalog,feb2011Catalog,2012Catalog,2013Catalog,2014Catalog}.  Kepler data in the prime mission were downlinked monthly but processed on a quarterly basis.  Transit searches and the associated planet candidate catalogs are, therefore, referred to by the quarters bracketing the data.  The most recent planet candidates were identified in a search of 12 quarters of data (Q1-Q12) where the first is only slightly longer than one month in duration (hence the 34.5 month time span).  

Previously detected candidates are reexamined as larger data volumes become available.  However, this does not occur with every catalog release.  Some of the candidates in the cumulative archive at the NEA were discovered with less than 34.5 months of data and have not yet been reexamined.  This non-uniformity will be resolved as Kepler completes its final search and vetting of the entire 17 quarters (48 months) of data acquired during its primary mission lifetime.  Kepler's planet candidate catalog is also known as the KOI (Kepler Object of Interest) Catalog.  However, KOIs also include events that are classified as false alarms or astrophysical false positives.  Only those flagged as planet candidates in the NEA cumulative catalog are shown in Figure~\ref{fig:discoveries}.

The catalogs contain the five parameters produced by fitting a limb-darkened Mandel \&Agol \cite{man02} model to the observed flux time series assuming zero eccentricity: the transit ephemeris (period and epoch), reduced radius (\rpl$/$\rstar), reduced semi-major axis ($d/$\rstar), and impact parameter. To first order, the reduced semi-major is equivalent to the ratio of the planet-star separation during transit to the stellar radius. Despite its name, it is equivalent to $a/$\rstar\ (where $a$ is the semi-major axis) only in the case of a zero eccentricity orbit.  

Planet properties are also tabulated in the discovery catalogs.  Planet radius, semi-major axis, and insolation flux are computed from light curve parameters and knowledge of the host star properties (effective temperature, surface gravity, mass, and radius).  The {\it Kepler Input Catalog} (KIC) \cite{kic} contains the properties of stars in the Kepler field of view derived from ground-based broad and narrow band photometry acquired before launch to support target selection.  However, the KIC contains known deficiencies and systematic errors making it unsuitable for computing accurate planet properties \cite{verner2011, gaidos2013, pinsonneault2012, mann2012, muirhead2012}.  

A Kepler working group provides incremental deliveries of updated properties of all stars observed by Kepler with the long-term goal of increasing accuracy and quantifying systematics.  Accuracy is required for characterizing individual planetary systems.  Also, an understanding of planetary populations via occurrence rate studies requires a homogeneous database of the properties of all observed stars.  Towards this aim, the working group coordinates campaigns and collates atmospheric properties (temperature, surface gravity, and metallicity) derived from different observational techniques (photometry, spectroscopy, and asteroseismology) which are then fit to a grid of stellar isochrones to determine fundamental properties like mass and radius.

The planet radii plotted in Figure~\ref{fig:discoveries} (and~\ref{fig:hzDiscoveries}) are not taken directly from the NEA cumulative table.   Rather, the planet radii (and ancillary properties like insolation flux) are recomputed using the modeled light curve parameters and the Q1-Q16 catalog of star properties (also available at the NEA), so called because it is used as input to the  Q1-Q16 pipeline run.  The provenance of all values in the Q1-Q16 star properties catalog are described by Huber et al. \cite{huber} as is the strategy for future updates to the catalog.  Published properties of confirmed planets are utilized where available.

Looking forward, there is one year of data left to analyze.  The Q1-Q16 pipeline run searched for statistically significant, transit-like signals, also called Threshold Crossing Events (TCEs).  Over 16,000 events were identified.  The Q1-Q16 TCE list is archived at the NEA and described in \cite{tenenbaum}.  The list contains previously discovered planets, false positives, and eclipsing binaries as well as numerous false alarms.  Dispositioning will occur after a vetting process using the validation tests described in \cite{wuDV}.  

Efforts to produce an updated catalog of planet candidates are underway and should be completed in mid-2014.  Hundreds of new discoveries are expected, including the first small planet candidates in the habitable zone of G-type stars.  Moreover, Kepler data are in the public domain thereby enabling many additional discoveries.  Both the scientific community \cite{ofir,petigura1} and citizen science efforts \cite{lintott,wang} have yielded new candidates and confirmed planets.  Interesting new niches of parameter space have been opened up thanks to such efforts.  Notables include the first seven-planet system, KOI-351 \cite{schmitt}, a planet in a quadruple star system \cite{schwamb}, and objects in ultra-short orbits \cite{sanchis}.

\section{Planets in the Habitable Zone} \label{sec:hz}

Kepler's objective is to determine the frequency of earth-size planets in the habitable zone (HZ) of sun-like stars.  Defined as the region where a rocky planet can maintain surface liquid water, the HZ is a useful starting point for identifying exoplanets that may have an atmospheric chemistry affected by carbon-based life \cite{kasting}.  As we broaden our perspective, we stretch and prod the HZ limits. Abe et al. \cite{abe} and Zsom et al. \cite{zsom} consider the extreme case of arid Dune-like planets.  LeConte et al. \cite{leconte} and Yang et al. \cite{yang} consider the effects of rotation.   And Lissauer \cite{lissauer} considers the desiccation of planetary bodies before their M-type host stars even settle onto the main sequence.  There may not be a simple evolutionary pathway that lands an exoplanet inside of a well-defined habitable zone.  Regardless, it is of interest to understand the prevalence of planets with properties similar to Earth.  For Kepler's exoplanets, comparisons with Earth are made considering size (radius) and orbital environment (period or semi-major axis) assuming we have knowledge of the host star properties.  The orbital environment can also be characterized by the irradiation, or insolation flux, defined as  $F=(^{R_\ast}/_{R_\odot})^2(^{T_\ast}/_{T_\odot})^4(^{a_\oplus}/_{a_p})^2$.  The insolation flux of each planet candidate is shown in Figure 2 where the y-axis is the effective temperature of the host star.

\begin{figure}
\includegraphics[width=0.5\textwidth]{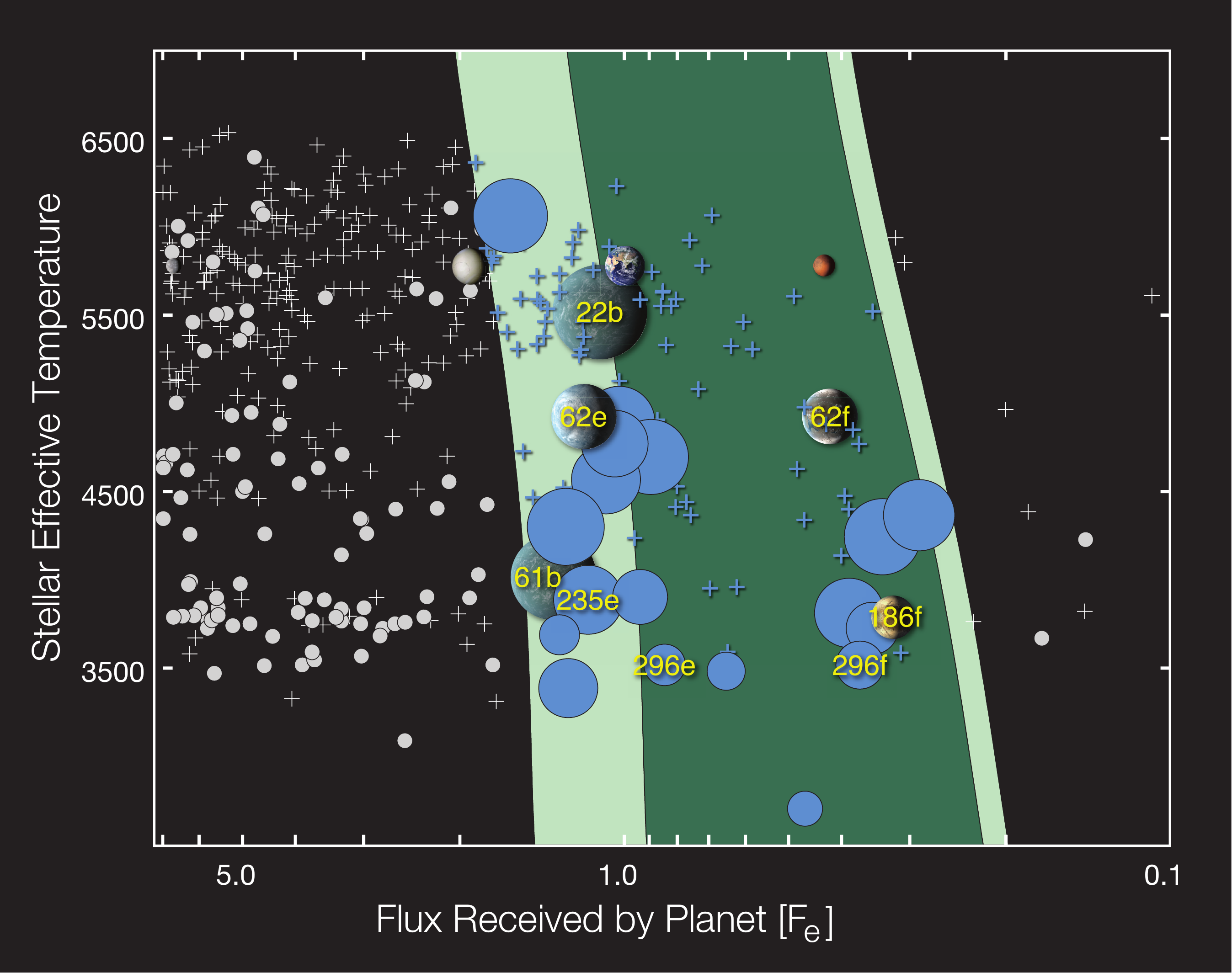}
\caption{Stellar effective temperature versus insolation (stellar flux at the semi-major axis) for Kepler exoplanets larger than 2 \rearth\ (plusses) and smaller than 2 \rearth\ (circles).  Symbols are colored blue if they lie within the habitable zone and are sized relative to the earth (shown as an image insert) if they represent a planet smaller than 2 \rearth.  The confirmed HZ exoplanets (Kepler-22b, Kepler-62e, Kepler-62f, Kepler-61b, and Kepler-186f) are displayed as artist's conceptions.}
\label{fig:hzDiscoveries}
\end{figure}

Two definitions of the habitable zone are included for reference in Figure~\ref{fig:hzDiscoveries}, both of which are taken from \cite{kopparapu_hz}.  The wider Habitable zone (light green) is based on the recent Venus and early Mars limits discussed therein and is referred to as the ``optimistic'' HZ.  The optimistic HZ does not extend all the way in to the Venusian orbit.  The Sun was $\sim$92\% as luminous a billion years ago at the epoch when Venus may have had liquid water on its surface.  The insolation intercepted by Venus during that epoch corresponds to the insolation at 0.75 AU in the present-day Solar System (1.78 F$_\oplus$).  Similarly, the outer edge of the optimistic HZ extends beyond the Martian orbit since the Sun was $\sim$75\% as luminous 3.8 billion years ago when Mars was thought to have liquid water.  The insolation intercepted by Mars at that epoch corresponds to the insolation at 1.77 AU in the present day Solar System (0.32 F$_\oplus$).   

The narrow HZ (dark green) is defined via climate models assuming an earth-mass planet with different CO$_2$ and H$_2$O compositions that take the planet to the two extremes.  These are the runaway greenhouse and maximum greenhouse limits \cite{kopparapu_hz} and are referred to as the ``conservative'' HZ.    According to these models, the highest flux a planet can receive while maintaining surface temperatures amenable to liquid water occurs for a water-saturated atmosphere.  The inner edge at 1.02 $F_\oplus$ corresponds to rapid water loss and hydrogen dissipation in a water-saturated atmosphere.  The outer edge at 0.35 F$_\oplus$ corresponds to the maximum possible greenhouse warming from a CO$_2$ dominated atmosphere.  Beyond the outer edge of this conservative habitable zone, models indicate that CO$_2$ begins to condense and lose its warming greenhouse properties.

The inner solar system planets line up horizontally in Figure~\ref{fig:hzDiscoveries}, with Mercury at the extreme left, Venus and Mars bracketing the optimistic HZ, and the Earth near the inner edge of the conservative HZ.  The HZ fluxes at the inner and outer edges have a slight dependence on the properties of the host star (note that the green shaded regions are not vertical bars).  The amount of radiation absorbed/reflected by the planet is wavelength dependent.  Therefore, the Bond albedo depends on the spectral energy distribution of the host star, and the limits are adjusted accordingly.

From the first three years of data (Q1-Q12), there are over one hundred candidates that have an insolation flux that falls within the optimistic habitable zone.  Of those, 21 are smaller than 2 $R_\oplus$.  These are shown as circles in Figure~\ref{fig:hzDiscoveries}. The symbols are sized in proportion to the Earth image to reflect their relative radii.  Five of the Kepler HZ discoveries are planets that have been statistically validated at the 99\% confidence level or higher:  Kepler-22b \cite{kepler22}, Kepler-61b \cite{kepler61}, Kepler-62 e \& f \cite{kepler62}, and Kepler-186f \cite{kepler186}, with radii of $2.38\pm0.13$, $2.15\pm0.13$, $1.61\pm0.05$, $1.41\pm0.07$ and $1.11\pm0.14$ $R_\oplus$, respectively.  These are represented by artist's conceptions, also scaled in size with respect to the Earth.  Kepler-235e and Kepler-296 e\&f are verified planets \cite{roweMultis} with uncertain properties.  Disparate star properties have been reported in the literature for Kepler-235.  The planet properties shown in Figure~\ref{fig:hzDiscoveries} are derived assuming a $0.48$ \rsun\ host star \cite{huber}.  Kepler-296 is a diluted (multiple star) system \cite{roweMultis}.  The properties of 296e and 296f shown here are derived assuming the planets orbit the primary star.  

Kepler's small HZ candidates orbit predominantly K and M-type main sequence stars -- perhaps not surprising given the fact that only 34.5 months of data were used to produce the sample of planet candidates shown in Figures~\ref{fig:discoveries} and~\ref{fig:hzDiscoveries}.  Habitable zone planets associated with G-type main sequence stars produce shallower transits, have longer orbital periods and, therefore, require more data for detection compared to those transiting cooler stars of comparable magnitude.  

Kepler was designed to achieve a 6.5-hour precision of 20 ppm or better for a 12th magnitude sun-like star.  The 20 ppm total allows for the detection of a 1.0 $R_\oplus$ planet with 4 transits, which in turn allows for the detection of Earth analogs in a 4-year mission.  The baseline noise budget for a G2-type main sequence star included a 10 ppm contribution for intrinsic stellar variability consistent with observations of the Sun \cite{jenkins02}.  However, the realized noise for 12th magnitude sun-like stars has a mode at 30 ppm due to a combination of unanticipated stellar variability and instrument noise \cite{gilliland}.  Both were a factor of two larger than expected and, when added to the shot noise, resulted in a total noise budget that is 50\% larger than anticipated.

An extended mission was awarded, but the loss of two of Kepler's reaction wheels degraded the pointing stability in the nominal field of view.  Very high pointing stability is required to achieve the photometric precision necessary to detect small HZ planets.  Consequently, the nominal mission ended with the loss of the second reaction wheel in May 2013.  Detection of sizable numbers of small HZ planets may require software solutions to reduce other noise contributions.  Numerous improvements to pipeline modules have been implemented, and a full reprocessing of the data is underway.

Considering the possibility of fewer detections than originally anticipated, it is critical to carefully quantify the reliability of the detections in hand.

\section{Catalog Reliability} \label{sec:reliability}

The Kepler discoveries are referred to as planet candidates until they are either dynamically confirmed or statistically validated (see below).  The former deals with follow-up observations and/or analyses that seek to identify dynamical evidence of an exoplanet (e.g. radial velocity or transit timing variations) whereas the latter deals with follow-up observations that seek to rule out scenarios produced by astrophysical signals that can mimic a planetary transit.  Potential sources of astrophysical false positives include:
\begin{enumerate}
\item Grazing eclipse of binary stars
\item Eclipse of a giant star by a main sequence star
\item Eclipse of an FGK-type main sequence star by a very late-type star or brown dwarf
\item Eclipse of a foreground or background binary near the target as projected on the sky
\item Eclipse of a binary physically associated with the target
\item Transiting planet orbiting a nearby (projected onto the sky) foreground or background star
\item Transiting planet orbiting a physical companion of the target star
\item Long-period, eccentric companion (star or giant planet) that yields only the secondary eclipse (or occultation). 
\end{enumerate}

Kepler's target stars are relatively well-characterized making it unlikely that an exoplanet transit will be confused by a main sequence star eclipsing a giant.  Moreover, Kepler's ultra-high precision photometry allows for statistical tests that eliminate many of the false positive scenarios that plague ground-based surveys.  For example, Kepler readily detects secondary eclipses of grazing and high-mass ratio eclipsing binaries.  Moreover, part-per-million differences between the eclipse depths of two nearly equal-mass stars are often discernible.  The statistical tests performed on the data to identify these tell-tale signs are described in \cite{wuDV}.

By design, Kepler's pointing stability is better than 0.003 arcsec on 15 minute timescales \cite{koch10}.  This allows us to measure relative star positions to milli-pixel precision \cite{bryson2013}.  The center of light distribution (photocenter) for a photometric aperture can be computed at each cadence producing a time series of row and column photocenter values with sub-milli-pixel precision on transit timescales \cite{batalhaDV}.  These time series contain information about the location of the source of the transit or eclipse event. However, dilution from multiple flux sources (known and unknown) in the aperture makes the interpretation difficult in some cases.  Alternatively, in-transit and out-of-transit pixel images can be used to construct difference images that provide direct information about the location of the transit (or eclipse) source \cite{bryson2013}.  Difference image analysis eliminates a large fraction of the false positive scenarios involving dilution from nearby targets.  

Follow-up observations further restrict the false positive parameter space.  Kepler has made it a priority to collect high resolution, high S$/$N spectra and high-contrast, high-spatial resolution imaging of as many of the planet-host stars as possible.  Difference image analysis rules out the presence of diluting stars outside of a spatial radius (typically about 2 arcsec, or a half of a pixel).  Adaptive optics or speckle imaging can tighten that radius to a fraction of an arcsecond, thereby significantly reducing the parameter space where false positives can lurk.  Bound stellar-mass companions with subarcsecond separation and flux greater than 1\% of the primary can be ruled out by spectroscopy \cite{kepler62}.  

Numerical simulations provide an estimate of the likelihood of remaining astrophysical false positive scenarios given the density of stars as a function of magnitude and galactic coordinates as well as the frequency of eclipsing binaries and transiting planets.  Morton \& Johnson \cite{mortonFP} compute the false positive probability (FPP) for each of the 1235 planet candidates reported in \cite{feb2011Catalog} and find the FPPs to be less than 10\% for nearly all candidates.  Empirical estimates are a mixed bag.  Santerne et al. \cite{santerneFP} perform radial velocity follow-up of 46 close-in giant planet candidates and estimate a 34.8\% false positive rate while D\'esert et al. \cite{desert} acquire Spitzer observations of  51 candidates (of primarily sub-Neptunian sizes) and identify only one false positive.

Fressin et al \cite{fressinFP} simulates the global population of astrophysical false positives that would be detectable in the observations of all target stars and would persist even after the careful vetting described above.  Two interesting results emerge.  Somewhat counterintuitively the highest false positive rates ($\sim$18\%) are found for the close-in giant planets which is qualitatively consistent with the empirical results of \cite{santerneFP}.  Secondly, the most common source of false positives mimicking small planets is a larger planet transiting an unseen physical companion or a background star.  Such scenarios were not considered in the Morton \& Johnson analysis.  Fressin et al. reports a $9.4\pm0.9$\% global false positive rate for the Q1-Q6 catalog \cite{2012Catalog}.  This value was revised upward \cite{santerne13} to $11.3\pm1.1$\% upon inclusion of secondary-only false positives.

Even if only 80-90\% of the detections are bona-fide planets, Kepler has quadrupled the number of exoplanets, providing a statistically significant and diverse population for studying demographics. 

\section{Planet Confirmation and Characterization} \label{sec:confirmations}

The confirmation and characterization of Kepler's exoplanet candidates contribute to planet population studies by increasing the reliability of the planet census and by offering an empirical ground-truth to estimates of false positive probabilities as previously discussed.  Just as important, however, is the information emerging about the distribution of planet densities.  With this information, we can estimate not only the occurrence rate of  ``earth-size'' planets in the habitable zone but also the occurrence rate of veritably rocky planets in the habitable zone.  

As of this writing, over 962 Kepler exoplanet candidates have been either dynamically confirmed or statistically validated.  High precision radial velocity follow-up has yielded $\sim$ 50 mass determinations from instruments scattered across the northern hemisphere,  including the SOPHIE spectrograph at the Observatoire de Haute-Provence \cite{santerne2011,bouchy2011,bonomo2012}, FIES on the Nordic Optical Telescope \cite{buchhave}, HRS on the Hobby-Ebberly Telescope \cite{endl}, HARPS-N on the Telescopio Nazionale Galileo \cite{hebrard}, and the HIRES spectrograph on Keck \cite{firstFive}.  Of special interest are the measurements for the sub-neptune size planets, particularly those that have densities indicative of a rocky composition: Kepler-10b \cite{kepler10b} and Kepler-78b \cite{kepler78b_1,kepler78b_2}.   A recent report on four years of strategic Keck observations \cite{marcyReview} has added another 6 candidate rocky planets to this roster.    

Dynamical confirmation is not limited to velocimetry measurements. Approximately half of Kepler's confirmations come from measurement of transit timing variations \cite{kepler9,ford_ttv2,steffen_ttv3,fabrycky_ttv4,steffen_ttv7,wu_ttv,xie_ttv1,xie_ttv2,ming_ttv}.  Anti-correlated timing variations exhibited by two planets in a system can place an upper limit on mass thereby supporting the planet interpretation.  In some cases, dynamical models of transit timing variations resulting from mutual planetary perturbations yield mass measurements.  Such measurements have been obtained for sub-neptune sized objects including five planets orbiting Kepler-11 \cite{kepler11_1,kepler11_2}, Kepler-20 b \& c \cite{kepler20_2}, Kepler-30b \cite{kepler30}, Kepler-18b \cite{kepler18}, Kepler-87c \cite{kepler87c}, Kepler-79 b \& c \cite{kepler79}, Kepler-36c, and its rocky neighbor, Kepler-36b \cite{kepler36}.   

Collectively, data on sub-Neptunian planets do not support a strict relation between mass and radius.  A power-law fit of mass versus radius for 63 exoplanets smaller than 4 \rearth\ has a reduced chi-square of 3.5 \cite{weiss2013}.  The large dispersion is indicative of a compositional diversity arising from the varied formation, migration, interaction, and irradiation pathways of planetary evolution.  Kepler-11d and Kepler-100b exemplify this diversity, having similar masses ($7.3\pm1.2$ and $7.3\pm3.2$ \mearth) but quite different radii ($3.12\pm0.07$ and $1.32\pm0.04$ \rearth).  Kepler-11d most likely contains a high H/He and/or ice envelope fraction ($\rho=1.28\pm0.20$ gcc) while Kepler-100b is consistent with an earth-like composition ($\rho=14.25\pm6.33$ gcc).  

Theoretical models of sub-neptune sized planets suggest that planetary radius changes very little with increasing mass for a given compositional mix \cite{lopez2013}.  The authors suggest that planetary radius is, to first order, a proxy for planetary composition.  However, the observational data serve as a caution.  Kepler-11b and Kepler-113b have nearly equal radii ($1.80\pm0.04$ and $1.82\pm0.05$ \rearth) yet  different masses ($1.9\pm1.2$ and $11.7\pm4.2$ \mearth) and densities ($1.72\pm1.08$ and $10.73\pm3.9$ gcc).  This occurs as well for planets in the same system. Kepler-138c and Kepler-138d, for example, have the same radius ($1.61\pm0.16$ \rearth) but different masses ($1.01^{+0.42}_{-0.34}$ and $3.83^{+1.51}_{-1.26}$  \mearth\, respectively) \cite{kipping}.

The fraction of planets of a given composition is likely to be a smooth function of planet size, implying no particular radius that marks a clean transition from rocky planets to those with H/He and/or ice envelopes.  There are hints, however, that most planets smaller than 1.5 \rearth\ are rocky while most planets larger than 2 \rearth\ have volatile-rich envelopes \cite{weiss2013}.  Moreover, planets larger than 3 \rearth\ are most often less dense than water, implying a higher hydrogen content in the atmosphere \cite{hadden}.  This suggests that the (somewhat arbitrary but commonly used) definition of ``earth-size'' (\rpl$ < 1.25 $\rearth) is in need of revision.

\section{Requirements for Reliable Planet Occurrence Rates} \label{sec:requirements}

Kepler's primary mission objective is to study exoplanet populations.  Of particular importance is the determination of $\eta_\oplus$  -- the frequency of earth-size habitable zone planets.  Though no discrimination by star type is captured in this definition, Kepler was designed with earth analogs in mind: earth-size planets in the HZ of G-type main sequence stars.  The determination of reliable planet occurrence rates requires:
\begin{enumerate}
\item Sensitivity to small HZ planets for sufficiently large numbers of G, K, and M stars
\item A uniform and reliable catalog of exoplanets with well-understood properties (radius, periods, etc.).
\item Knowledge of Kepler's detection efficiency as a function of both planet and star properties.
\item Knowledge of the catalog reliability as a function of both planet and star properties.
\item Well-documented and accessible data products for future archive studies.
\end{enumerate}
As previously mentioned, sensitivity to earth analogs orbiting G-type stars is a challenge that is being tackled with improvements to software.  Planet properties depend on knowledge of star properties, and work is underway to construct a catalog of accurate properties and characterize systematics.  Catalog uniformity is achieved by removing human subjectivity from the discovery process.  Each time a new planet candidate catalog is generated, there are fewer manual processes thereby improving uniformity. A machine-learning algorithm approach based on a random-forest classifier is simultaneously being developed and may eventually replace the manual processes altogether \cite{autoVetter}.  

The detection efficiency is computed by injecting artificial transits at both the pixel level and the flux level.  Artificial transits are propagated through the system from pixels to planets to quantify the completeness end-to-end.  Tests on the back end of the pipeline (pixel calibration,  aperture photometry, systematic error correction, harmonic variability removal) demonstrate a 98\% fidelity in preserving the signal-to-noise ratio of a single transit \cite{christiansen}.  The tests will be repeated with longer data volumes.  Tests on the front end of the pipeline (whitening filters, signal detection, and vetting) are in progress. The false positive probability is computed for every planet candidate as described above \cite{morton12}, yielding a quantitative measure of catalog reliability.

Knowledge of the statistics of multiple star systems is crucial to several key studies.  They are used to construct priors for statistical validation, for computing the FPP for planet-hosting stars, and for estimating the catalog reliability.  They are also required for computing Kepler's detection efficiency.  The probability of detecting a planet of a given size and orbital period around a star is degraded in the presence of flux dilution from unresolved nearby stars (either bound or line-of-sight).  Contaminating flux causes transits to appear shallower.  We do not know a priori which stars have such dilution.  However, the effect on occurrence rates can be quantified via numerical simulation based on multiple star statistics from Kepler  \cite{eb1,eb2} and other \cite{raghavan} surveys.    

It is important to note that the sample of stars observed by Kepler is not representative of the Galactic population \cite{batalha_tm}.  Exoplanet occurrence rates must be broken out by star type in order to reconstruct a volume-limited representation of planetary populations in the Galaxy.  Finally, if future missions need to know how deeply they must probe before happening on a potentially habitable terrestrial planet, we must consider how the Kepler planets cluster into multi-planet systems and compute the fraction of stars with planets in addition to the average number of planets per star.

\section{Estimates of Planet Occurrence Rates}\label{sec:statistics}

There has yet to be a study that addresses all of the requirements described above using all of the available data.  Nevertheless, numerous population estimates have been reported in the literature and patterns are beginning to emerge.  The most dramatic is the sharp rise in the (log) radius distribution for planets smaller than about 3 times the size of earth \cite{howard, fressinFP}.  

\begin{figure*}
\includegraphics[width=1.0\textwidth]{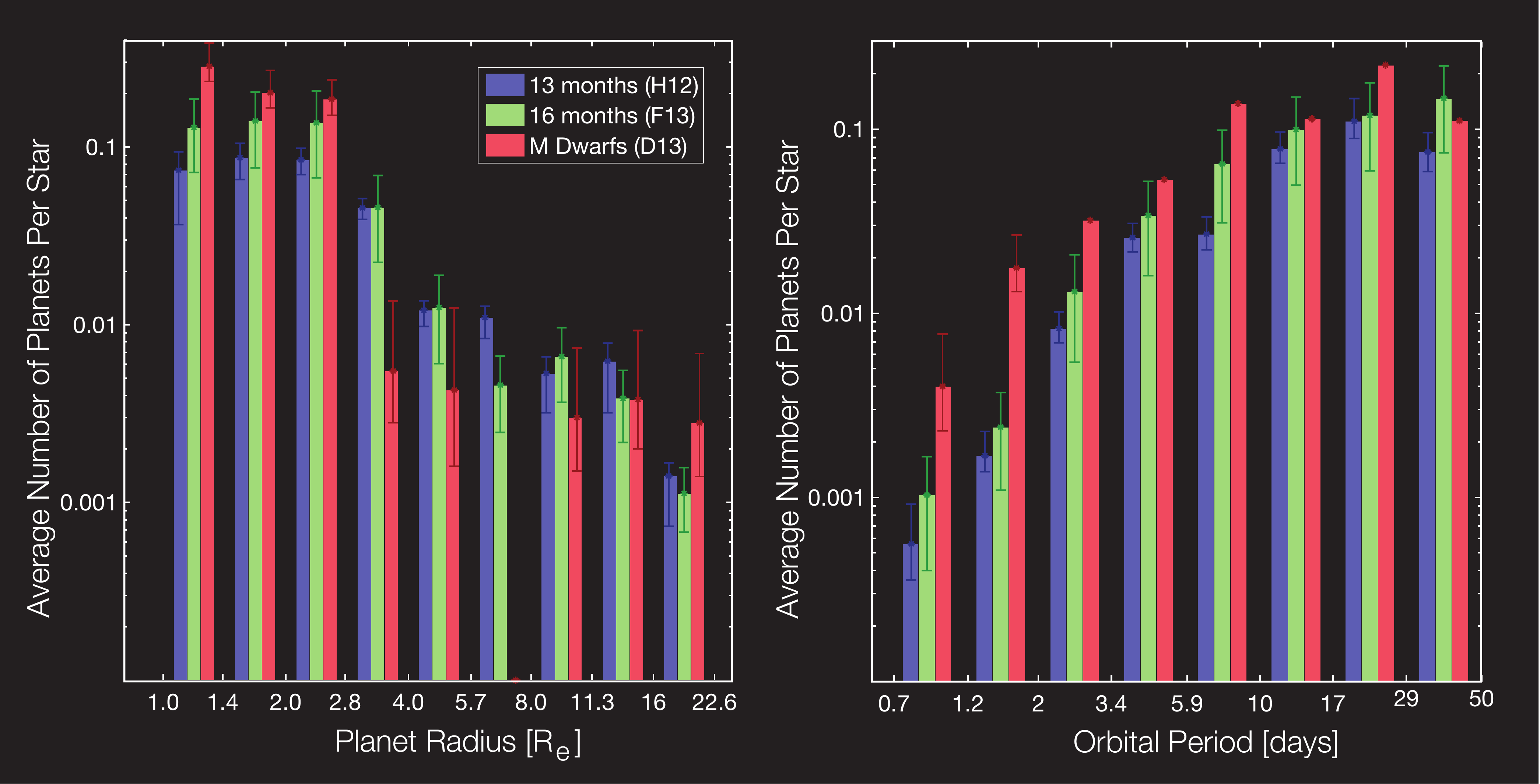}
\caption{The radius distribution (left) and period distribution (right) of planet occurrence rates expressed as the average number of planets per star.  The distributions have been marginalized over periods between 0.68 and 50 days (radius distribution) and radii between 0.5 and 22.6 \rearth\ (period distribution).  H12 refers to \cite{howard}; F13 refers to \cite{fressinFP}; D13 refers to \cite{dressing}.}
\label{fig:statistics}
\end{figure*}

Figure~\ref{fig:statistics} (left) shows the planet occurrence rate distribution marginalized over periods less than 50 days reported by independent teams (0.68 to 50 days being the common domain).  A power law distribution would be a straight line on this logarithmic display.  Close-in giants are orders of magnitude less common than planets smaller than Neptune.   However, a power law increase toward smaller sizes is not observed.  The distribution flattens out for planets smaller than 2 \rearth. This may be an artifact of catalog incompleteness for the smallest planets, especially at longer orbital periods.   

Marginalizing over radius (0.5 - 22.6 \rearth), we observe a power law increase in occurrence rate as a function of (log) period up to approximately 10 days.  At longer orbital periods, the distribution flattens (Figure~\ref{fig:statistics}, right).  The trend can be explored with a larger sample that includes longer period planets.   The flat distribution persists out to $~\sim 250$ days \cite{dong} at least for planets smaller than neptune.  The giants, however, appear to be gaining ground, slowly increasing in frequency (cf. Figure 7 from \cite{dong}) -- a trend that is consistent with doppler surveys \cite{udry} and predicted by core-accretion models \cite{idalin}.

The habitable zone of M-type dwarfs corresponds to orbital periods of a few weeks to a few months.  Kepler's current planet catalog is sufficient for addressing statistics of HZ exoplanets orbiting M stars.  The results indicate that the average number of small (0.5 -1.4 \rearth) HZ (optimistic) planets per M-type main sequence star is approximately 0.5 \cite{dressing,kopparapu_M,gaidos_hz}.   An estimate of HZ occurrence rates for G and K stars has been made via extrapolation to longer orbital periods \cite{petigura2}.   An independent planet detection pipeline was applied to a sample of G and K stars observed by Kepler, and the survey completeness was quantified via signal injection.  An occurrence rate of $11 \pm 4$\% was recovered for 1-2 \rearth\ planets receiving insolation fluxes of 1-4 F$_\oplus$.  

Assuming the true occurrence rate distribution is approximately constant in (log) period for P$>10$ days and in (log) radius for \rpl\ $< 2.8$ \rearth, the planet occurrence over any interval within that domain is proportional to the logarithmic area bounded by the interval.  For a homogeneous star sample, a distribution that is constant in (log) period will, to first order, be constant in (log) insolation flux.  An orbital period of 10 days corresponds to an insolation flux  of $\sim100$ F$_\oplus$ for a sun-like star ($\sim20$ F$_\oplus$ for a late K).  

Under these assumptions, the reported occurrence rate of $11 \pm 4$\% can be scaled for small planets (1-1.4 \rearth) in the optimistic HZ (0.27-1.70 F$_\oplus$ for a K0-type main sequence star).  This yields an occurrence rate of $7 \pm 3$\%.  If we assume that the (log) radius distribution remains constant down to $0.5$ \rpl, we can estimate the occurrence rate for an interval comparable to that of the M dwarf calculations (0.5 - 1.4 \rearth\, optimistic HZ).  The G and K occurrence rate for this interval is $22 \pm 8$\%.  At first glance, planets orbiting in the HZ of G and K-type stars are less common than those orbiting M-type stars.  We must proceed cautiously, however, since the results are based on extrapolation to longer periods to account for very high incompleteness.

Collectively, the statistics emerging from the Kepler data suggest that every late-type main sequence star has at least one planet (of any size), that one in six has an earth-size planet within a mercury-like orbit, and that small HZ planets around M dwarfs abound.  Already, the Kepler data suggest that a potentially habitable planet resides within 5 parsecs at the 95\% confidence level.

\section{Summary}

Our blinders to small planets have been lifted, and the exoplanet landscape looks dramatically different than it did before the launch of NASA's Kepler Mission.  A picture is forming in which small planets abound and close-in giants are few, in which the habitable zones of cool stars are heavily populated with terrestrial planets and the diversity of systems challenges preconceived ideas.  The picture will continue to evolve over the next few years as we analyze the remaining data, refine the sample, and quantify the observational biases.  Characterization instruments will continue to gain sensitivity ensuring that Kepler's exoplanet discoveries will be studied for years to come.  Although Kepler's primary data collection has officially ended, the most significant discovery and analysis phase is underway, enabling the long-term goal of exoplanet exploration: the search for habitable environments and life beyond the solar system.

\begin{acknowledgments}
The significant science results based on Kepler data described in this volume would not have been possible without the decades of persistence, hard work, creativity, and expertise of William Borucki (Principal Investigator),  Dave Koch (Deputy Principal Investigator), Jon Jenkins (Analysis Lead), and Doug Caldwell (Instrument Scientist).  Kepler's co-Investigators, Science Working Group members, Follow-up Observers, and Participating Scientists (PSP) provide critical expertise and analyses that help Kepler meet its baseline objectives.  The Science Operations Center (SOC) provides the software pipeline for constructing light curves and identifying planet candidates.  The Science Office, led by Michael Haas, produces the well-vetted and highly reliable catalogs that Kepler's planet occurrence rates are derived from.  This research has made use of the NASA Exoplanet Archive, which is operated by the California Institute of Technology, under contract with the National Aeronautics and Space Administration under the Exoplanet Exploration Program.   Kepler was competitively selected as the tenth Discovery mission. Funding for this mission is provided by NASAÕs Science Mission Directorate.
\end{acknowledgments}





\end{article}



\end{document}